\documentclass[twocolumn,showpacs,preprintnumbers,amsmath,amssymb,prl]{revtex4}

\usepackage{graphicx}
\usepackage{dcolumn}
\usepackage{bm}

\providecommand{\mcal}{\ensuremath{\mathcal}}
\newcommand{\transker}[1][\opr{T}]{\ensuremath{\left.\left<q+\frac{v}{2}\right|#1\left|q-\frac{v}{2}\right>\right.}}
\newcommand{\opr}[1]{\ensuremath{\mathbf{\mathsf{#1}}}}

\newcommand{\abs}[1]{\ensuremath{\left|#1\right|}}

\newcommand{\kernel}[1]{\ensuremath{\left<q\left|#1\right|q'\right>}}

\newcommand{\kernelq}[3]{\ensuremath{\left<#1\left|#2\right|#3\right>}}

\newcommand{\dirac}[2]{\ensuremath{\left<#1\left|\right.#2\right>}}

\newcommand{\sign}[1]{\ensuremath{\mbox{sgn}\left(#1\right)}}

\newcommand{\expo}[1]{\ensuremath{\mbox{e}^{#1}}}

\begin{document}

\title{\bf Theory of Confined Quantum Time of Arrivals}

\author{Eric A. Galapon}
\email{eric.galapon@up.edu.ph}
\affiliation{Theoretical Physics Group, National Institute of Physics, University of the Philippines, Diliman Quezon City, 1101 Philippines}
\affiliation{Theoretical Physics, The University of the Basque Country, Apdo. 
644, 48080 Bilbao, Spain}
\affiliation{Chemical Physics, The University of the Basque Country, Apdo. 
644, 48080 Bilbao, Spain}

\date{\today}

\begin{abstract}
We extend the concept of confined quantum time of arrival operators, first developed for the free particle [E.A. Galapon, R. Caballar, R. Bahague {\it Phys. Rev. Let.} {\bf 93} 180406 (2004)], to arbitrary potentials. 
\end{abstract}


\maketitle

Quite recently the quantum time problem \cite{gen} has reached a new episode, starting from the elucidation that Pauli's theorem \cite{pauli} does not hold in Hilbert space \cite{galapon1}, followed by the discovery that the classical free time of arrival admits self-adjoint and canonical quantization for a spatially confined particle \cite{galapon0}. There it is shown that formulating the quantum-time-of-arrival-problem in a segment of the real line suggests rephrasing the problem to finding a complete set of states that unitarily arrive at a given point at a definite time---states in which the events of the centroid being at the origin and the position distribution width being minimum occur at the same instant of time. Specifically, it is demonstrated that, for a spatially confined particle, the problem admits a solution in the form of an eigenvalue problem of a compact and self-adjoint time of arrival operator derived by a quantization of the classical time of arrival. In this Letter, we attempt to extend the results in \cite{galapon0} and develop a formalism in constructing confined time of arrival operators for arbitrary potentials. 

In \cite{galapon0} the well-known non-self-adjointness of the free time of arrival operator $\opr{T}=-\mu (\opr{qp^{-1}+p^{-1}q})2^{-1}$ in the Hilbert space $\mcal{H}_{\infty}=L^2(-\infty,\infty)$ has been dealt with by spatial confinement of the particle. That is by projecting the operator $\opr{T}$ in the Hilbert space $\mcal{H}_l=L^2[-l,l]$, with the projected operator still satisfying conjugacy with the Hamiltonian in a closed subspace of $\mcal{H}_l$.
This has led to a class of self-adjoint and compact time of arrival operators, $\opr{T}_{\gamma}$, in the form of a Fredholm integral operator in position representation. There it has been demonstrated that the eigenfunctions of these operators unitarilly arrive at the origin at their respective eigenvalues. Now if the particle is subject to some potential $V(q)$, can the first time of arrival states at the origin and their respective arrival times be solved also in the form of an eigenvalue problem of an appropriately constructed confined time of arrival operator?

To answer this, we have to address first the question of how one constructs confined quantum time of arrival operators (at the origin) for arbitrary potentials.  For a given Hamiltonian $H(q,p)$, one may propose at quantizing the classical time of arrival $T(q,p)=-\sign{p} \sqrt{\mu/2}\int_0^q \left(H(q,p)-V(q')\right)^{-\frac{1}{2}} dq'$ to give a formal time of arrival operator that will be subsequently projected to the the Hilbert space $\mcal{H}_l$. However, naive quantization of $T(q,p)$ has two problems. First, it is known that there exists an obstruction to quantization in Euclidean space \cite{got11}; that is, the classical Possion brackets do not generally carry over to the required commutator relation. In general, this obstruction frustrates any effort at finding a quantization of $T(q,p)$ for arbitrary Hamiltonian $H(q,p)$ such that the classical canonical relation $\left\{H,T\right\}_{PB}=1$ goes over to the quantum canonical relation $\left[\opr{H},\opr{T}\right]=i\hbar\opr{I}$ \cite{galapon3}. Second, even when we are contented with quantization without the required algebra of observables, the quantity $T(q,p)$ is generally not everywhere real valued and can be multiple valued even when real. This makes its quantization unclear and its interpretation ambiguous. In extending our result in \cite{galapon0} to arbitrary potentials, we require consistency with the canonical commutation relation and demand unambiguous interpretation of the resulting operators. 

The consistency requirement behooves us to find the quantum image outside the framework of quantization, especially for those that cannot be quantized consistently. Reference-\cite{galapon3} provides us with the principles to go around the existing obstruction in Euclidean space, at least for the problem at hand. The basic idea in \cite{galapon3} is to presuppose that a given class of quantum observables, such as quantum time of arrivals, has an identifying set of properties; or, equivalently, that the observables comprising the class share a common set of properties. The problem now is, first, to find this shared set of properties; and then, second, on the basis of these properties, find all observables of the class. It is further presupposed in \cite{galapon3} that the first can be accomplished by studying a few known observables of the class; and then the second is accomplished by employing a transfer principle, i.e. by transferring the extracted shared properties from the known elements without discrimination to the rest of the class.

But how are we to determine the few observables of the class of observables to start with? We note that a quantization exists that preserves the classical algebra for a small class of observables despite the existence of obstruction. We can use this fact as a tool in determining the few observables to start with. Given a class $\mcal{C}$ of classical observables, we divide $\mcal{C}$ in two parts: The non-obstructed class, $\mcal{C}_N$, consisting of those observables that can be quantized such that the Possion-bracket-commutator correspondence, $\{ , \}_{P.B.}\mapsto \frac{1}{i\hbar}[ , ]$, is satisfied; and the obstructed class, $\mcal{C}_O$, consisting of those observables that can not be quantized to satisfy the Possion-bracket-commutator correspondence. Now we choose the non-obstructed class to start with. The quantum image of $\mathcal{C}_N$, $\hat{\mathcal{C}}_N$, is then found by quantization. We then identify the shared set of properties of those in $\hat{\mathcal{C}}_N$ that can be transferred to the quantum image of the obstructed class $\mathcal{C}_O$, $\hat{\mathcal{C}}_O$. But what are those shared properties? Those are the properties that uniquely identify the observables in $\hat{\mathcal{C}}_N$; and, at once, the properties that ensure that the observables in $\hat{\mathcal{C}}_O$ satisfy the correspondence $\{ , \}_{P.B.}\mapsto \frac{1}{i\hbar}[ , ]$ and the correspondence $\hat{\mathcal{C}}_O \mapsto \mathcal{C}_O$ in the classical limit. By virtue of the transfer principle, the observables in $\hat{\mathcal{C}}_O$ are then found by imposing the identified shared properties. In our language in \cite{galapon3}, the observables in $\hat{\mathcal{C}}_O$ obtained by the method just described are the supraquantizations of the classical observables in $\mathcal{C}_O$.

Now we proceed in developing a theory of confined quantum time of arrivals using the above ideas. With our intention to employ quantization to determine the shared properties of CTOA-operators, we now address the unambiguous quantization of the classical time of arrival. The idea is not to quantize $T(q,p)$ but its expansion in the neighborhood of the origin in the form $T(q,p)=\sum_{k=0}^{\infty}(-1)^k T_k(q,p)$, where the $T_k$'s are determined recursively through $T_0(q,p)=-\mu qp^{-1}$, and $T_k (q,p)=-\mu p^{-1}\int_{0}^q (\partial_{q'} V) (\partial_{p} T_{k-1})\,dq'$, in which $V$ is assumed continuous in every neighborhood of the origin. We referred to this form of $T(q,p)$ as the local time of arrival (LTOA) in \cite{galapon3}. For sufficiently small neighborhoods, the LTOA is a first time of arrival at the origin; and quantization of the LTOA where the required algebra is satisfied should give us a first time of arrival operator, as our numerical results below suggest. 

But how should we quantize the LTOA? We impose metaplectic covariance, hermicity of the resulting operator, and preservation of the classical symmetry in the quantum regime. Of all quantizations satisfying the first criterion, only the Weyl quantization satisfies the other two. Generally the LTOA is going to be in the form $T(q,p)=\sum_{m,n\geq 0}\alpha_{m,n}\, q^n p^{-m}$, so that its quantization proceeds by the replacement $q^n p^{-m} \mapsto \opr{T}_{m,n} = 2^{-n}\sum_{j=0}^{n} {n \choose j} \opr{q}^j \opr{p}^{-m} \opr{q}^{n-j}$. With respect to Weyl quantization, the non-obstructed class $C_N$ of the LTOA-observables consists of linear systems, and the obstructed $C_O$ class consists of the non-linear systems \cite{galapon3}. We then proceed by getting the quantum image, $\hat{\mathcal{C}}_N$, of $\mathcal{C}_N$ by means of Weyl quantization. The CTOA-operators, $\hat{\mathcal{C}}_N^{(c)}$, corresponding to $\mathcal{C}_N$ are then obtained by projecting the entire $\hat{\mathcal{C}}_N$ in the Hilbert space $\mathcal{H}_l$. Then the CTOA-operators , $\hat{\mathcal{C}}_O^{(c)}$, corresponding to $\mathcal{C}_O$ are found from $\hat{\mathcal{C}}_N^{(c)}$ by virtue of the transfer principle.

First let us demonstrate our idea through the harmonic oscillator; its Hamiltonian is $H(q,p)=(2\mu)^{-1}p^2+ \mu \omega^2 2^{-1} q^2$. This is a linear system, and it belongs to the non-obstructed class $\mcal{C}_N$. The classical time of arrival at the origin is given by $T_0(q,p)=-{\omega}^{-1}\tan^{-1}\!\left(\mu\omega q p^{-1}\right)$, which is multiple valued. As such its quantization is not clear. Now its local time of arrival at the origin is given by $T(q,p)=-\sum_{k=0}^{\infty}\frac{(-1)^k}{2k+1}\mu^{2k+1}\omega^{2k} q^{2k+1} p^{-(2k+1)}$.
And Weyl quantization of this leads to the operator $\opr{T}=\sum_{k=0}^{\infty}\frac{(-1)^k}{2k+1}\mu^{2k+1}\omega^{2k}\opr{T}_{2k+1,2k+1}$. Using the canonical commutation relation $\left[\opr{q},\opr{p}\right]=i\hbar \opr{I}$, formally one can show that $\left[\opr{H},\opr{T}\right]=i\hbar\opr{I}$, demonstrating the fact that Weyl quantization is not obstructed for linear systems. 

Now we show how the corresponding CTOA-operator is constructed by projecting $\opr{T}$ in $\mcal{H}_l=L^2[-l,l]$. Physically the projection is by spatial confinement of the particle in the interval $[-l,l]$ under the condition that the evolution of the system is generated by a purely kinetic self-adjoint Hamiltonian---i.e. $\opr{H}=(2\mu)^{-1} \opr{p}^2$, where $\opr{p}$ is a self-adjoint momentum operator---when the potential vanishes. This requirement projects the momentum operator $\opr{p}$ into the ring of momentum operators $\left\{\opr{p_{\gamma}}=-i\hbar \partial_{q},\, \abs{\gamma}<\pi\right\}$, with $\opr{p}_{\gamma}$ having the  domain consisting of those vectors $\phi(q)$ in $\mcal{H}_l$ with square integrable first derivatives, and satisfying the boundary condition  $\phi(-l)=\expo{-2i\gamma}\phi(l)$. Since $\opr{T}$ depends on the momentum operator, the projection of $\opr{T}$ in $\mcal{H}_l$  is a family of operators $\left\{\opr{T}_{\gamma}\right\}$, with each $\opr{T}_{\gamma}$ corresponding to the momentum $\opr{p}_{\gamma}$. 

The explicit form of $\opr{T}_{\gamma}$ is found by first writing the explicit forms of the operators $\opr{T}_{m,n}^{\gamma}$ in $\mcal{H}_l$ for a given $\gamma$, in particular their kernels in coordinate representation. In the development of the CTOA-operators only the cases where $m(s)=2s+1$, $s=0, 1, \dots$, are relevant. For these cases, the kernels are given by
\begin{eqnarray}
\kernel{\opr{T}^{\gamma\neq 0}_{2s+1,n}}\!&=&\!\frac{1}{2 \sin\gamma} \frac{(q+a')^n}{2^n}\frac{(-1)^s(q-q')^{2s}}{\hbar^{2s+1} (2s)!}\nonumber\\
& &\! \times\!\!\left(e^{i\gamma}H(q-q')+e^{-i\gamma}H(q'-q)\right),\label{an1}
\end{eqnarray}
\vspace{-0.2in}
\begin{eqnarray}
\kernel{\opr{T}_{2s+1,n}^0}&=& \frac{i}{l}\frac{(-1)^s}{2 \hbar^{2s+1}}\frac{(q+q')^n}{2^n}\frac{(q-q')^{2s}}{(2s)!}\mbox{sgn}\!(q-q')\nonumber\\
& &-\frac{i}{l}\frac{(-1)^s}{2 \hbar^{2s+1}}\frac{(q+q')^n}{2^n}\frac{(q-q')^{2s+1}}{(2s+1)!}.\label{an2}
\end{eqnarray}

Substituting $\opr{T}_{m,n}^{\gamma}$ back in $\opr{T}$ gives a compact and self-adjoint operator $\opr{T}_{\gamma}$ for every $\gamma$ in the form of a Fredholm integral operator $(\opr{T}_{\gamma}\varphi)\!(q)=\int_{-l}^l \kernel{\opr{T}_{\gamma}}\varphi(q')dq'$ in position representation. The respective kernels for the non-periodic and periodic cases can be explicitly evaluated using equations-(\ref{an1}) and-(\ref{an2})  to give
\begin{eqnarray}
\kernel{\opr{T}_{\gamma\neq 0}}&=&-\frac{1}{2\omega\,\sin\gamma}\frac{\sinh\left(\frac{\mu\omega}{2\hbar} (q^2-q'^2 \right)}{(q-q')}\nonumber\\
& & \times\left(e^{i\gamma}H(q-q')+e^{-i\gamma}H(q'-q)\right)\nonumber
\end{eqnarray}
\vspace{-0.2in}
\begin{eqnarray}
	\kernel{\opr{T}_{0}}&=&\frac{1}{2 i \omega} \frac{\sinh\left(\frac{\mu\omega}{2\hbar}          		(q^2-q'^2)\right)}{(q-q')}\mbox{sgn}(q-q')\nonumber\\
	& & - \frac{1}{2 i l\omega} 			\mbox{shi}\left(\frac{\mu\omega}{2\hbar}(q^2-q'^2)\right)\nonumber
\end{eqnarray}
where $\mbox{shi}(x)$ is the $\sinh$-integral function and $H(x)$ is the Heaviside function. Both kernels are essentially bounded everywhere in the plane $[-l,l]\times [-l,l]$. The singularities at the diagonal are removable, hence for every $\gamma$, $\kernel{\opr{T}_{\gamma}}$ is square integrable. Moreover, the kernels are symmetric. Thus for all $\gamma$, the operator $\opr{T}_{\gamma}$ is self-adjoint and compact. Notice that while the classical LTOA is valid only in a small neighborhood of the origin, the projection of its quantization is well-defined in $\mcal{H}_l$ for any finite value of $l$. 

The observables of $\hat{\mathcal{C}}_N^{(c)}$ corresponding to the non-obstructed class $\mathcal{C}_N$, systems whose potentials are $V(q)=c+ a q + \frac{1}{2} b q^2$, can now be constructed similarly. The LTOA at the origin can be derived using the recurrence relation given earlier. And Weyl quantization of this local form gives a formal operator $\opr{T}$ whose projection in $\mcal{H}_l$ is the family of integral operators $\left\{(\opr{T}_{\gamma}\varphi)\!(q)=\int_{-l}^l \kernel{\opr{T}_{\gamma}}\varphi(q')dq'\right\}$. The respective kernels for non-periodic and periodic boundary conditions are given by
\begin{equation}\label{timekernel}
	\kernel{\opr{T}_{\gamma\neq 0}}\!=\!-\mu\frac{T(q,q')}{\hbar\sin\gamma} \left(e^{i\gamma}H(q-q')\!+\!           			e^{-i\gamma}H(q'-q)\right);
\end{equation}
\vspace{-0.2in}
\begin{equation}	\label{periodic}
\kernel{\opr{T}_{0}}\!\!=\!\!\frac{\mu}{i\hbar}T(q,q') \mbox{sgn}(q\!-\!q')
\!\!-\!\!\frac{\mu}{i l\hbar}\!\!\! \int_0^{(q-q')}\!\!\!\!\!\!\!T\!\!\left(\!\!\frac{q\!+\!q'}{2},s'\!\!\right)\! ds'
\end{equation}
where $T(q,q')$ is a function characteristic and unique to the system, whose exact form, which follows from equations-(\ref{an1}) and-(\ref{an2}), we need not give it here. $T(q,q')$ is symmetric, i.e. $T(q,q')=T(q',q)$, real valued, and analytic everywhere in the plane $[-l,l]\times[-l,l]$.
These operators are compact and self-adjoint. Inspecting equations-(\ref{timekernel})and-(\ref{periodic}), we can identify that the functional forms of the kernels are shared properties of the observables of the non-obstructed $\hat{\mathcal{C}}_N^{(c)}$.

For non-linear systems, Weyl quantization of the LTOA at the origin does not give an operator conjugate with the Hamiltonian. This is the obstruction to quantization at work. Assuming a transfer principle, we require that the confined time of arrival operators, $\hat{\mathcal{C}}_O^{(c)}$, for the obstructed class $\mathcal{C}_O$ to be given by the family of integral operators $\left\{(\opr{T}_{\gamma}\varphi)\!(q)=\int_{-l}^l \kernel{\opr{T}_{\gamma}}\varphi(q')dq'\right\}$ whose kernels are likewise given by equations-(\ref{timekernel}) and-(\ref{periodic}), with $T(q,q')$ to be determined. Assuming in the mean time that $V(q)$ is $C^{\infty}(-l,l)$, let $\Phi^{\times}_l\supset\mcal{H}_l\supset\Phi_l$ be the rigging of the Hilbert space $\mcal{H}_l$, where $\Phi_l$ is the space of infinitely differentiable functions which vanish at the boundaries together with their derivatives of all orders. 
In the rigging $\Phi^{\times}_l\supset\mcal{H}_l\supset\Phi_l$, the conjugacy relation $\left[\opr{H},\opr{T}\right]=i\hbar \opr{I}$ satisfied by the non-obstructed CTOA-operators with their respective Hamiltonians, in particular for $\gamma\neq 0$, translates to the canonical commutation relation (CCR)
$\kernelq{\tilde{\varphi}}{\left[\opr{H}^{\times}_{\gamma},\opr{T}_{\gamma}\right]}{\varphi}=i\hbar\dirac{\tilde{\varphi}}{\varphi}$
for all $\tilde{\varphi}$ and $\varphi$ in $\Phi_l$, where $\opr{H}_{\gamma}^{\times}$ is the Rigged Hilbert space extension of the system Hamiltonian $\opr{H}_{\gamma}=(2\mu)^{-1}\opr{p}_{\gamma}^2+V(q)$. This is the {\it basic requirement} we impose on all CTOA-operators. 

Substituting equation (\ref{timekernel}) back into the left hand side of $\kernelq{\tilde{\varphi}}{\left[\opr{H}^{\times}_{\gamma},\opr{T}_{\gamma}\right]}{\varphi}=i\hbar\dirac{\tilde{\varphi}}{\varphi}$, we find that $\opr{H}^{\times}_{\gamma}$ and $\opr{T}_{\gamma}$ satisfy the CCR if and only if $T(q,q')$ satisfies
\begin{equation} \label{kerneldiff}
 -\frac{\hbar^2}{2\mu}\frac{\partial^2 T(q,q')}{\partial q^2}+\frac{\hbar^2}{2\mu} \frac{\partial^2 T(q,q')}{\partial {q'}^2}+ \left(V(q)\!-\!V(q')\right)T(q,q')\!\!=\!\!0
\end{equation}
\begin{equation}
\frac{d T(q,q)}{dq} + \frac{\partial T}{\partial q}(q',q') + 	\frac{\partial T}{\partial q'}(q,q)=1.\label{kwe}
\end{equation}
It can be shown that $T(q,q')$ for the non-obstructed class, which is derived via quantization of the LTOA, satisfies equations-(\ref{kerneldiff}) and-(\ref{kwe}). Now equation-(\ref{kwe}) defines a family of operators canonically conjugate to the extended Hamiltonian in the sense required by the canonical commutation relation. We fix $T(q,q')$ by identifying the condition that uniquely identifies the kernels for the linear systems. By transforming equation-(\ref{kerneldiff}) in its canonical form, one can show that the boundary conditions
\begin{equation}\label{bc}
		T(q,q)=\frac{q}{2},\;\;	T(q,-q)=0
\end{equation}
uniquely identifies the kernels for the non-obstructed class. By virtue of the transfer principle we require that the kernels of the CTOA-operators for the obstructed class are given by equations-(\ref{timekernel}) and-(\ref{periodic}), with $T(q,q')$ having the properties of those of the unobstructed class and determined by solving equation-(\ref{kerneldiff}) subject to the boundary conditions-(\ref{bc}). By a proper rigging of $\mcal{H}_l$, we can lift the condition on $V(q)$ and extend the same transfer principle. And we have accomplished the supraquantization of the CTOA-operators for the obstructed class.

The CTOA-operators for linear systems are projections of the quantizations of the local time of arrival operator so that they have the correct classical limit. How about the CTOA-operators for the non-linear systems? The kernels of the CTOA-operators goes over to the kernel of the time of arrival operator in the real line constructed without quantization in \cite{galapon3} in the limit as $l$ goes to infinity---the kernel $\kernel{\opr{T}}=-i\mu\hbar^{-1} T(q,q')\, \mbox{sgn}(q-q')$ \footnote{This is obvious for equation-(\ref{periodic}). For equation-(\ref{timekernel}), the limit $l\rightarrow \infty$ is equivalent to the limit $\gamma\rightarrow 0$. That is at infinity all wavefunctions must vanish so that the boundary condition becomes necessarily periodic.}. There it is shown that the Weyl-Wigner transform of this kernel, i.e.
$\mathcal{T}_{\hbar}(q,p)=2\pi\!\! \int_{-\infty}^{\infty}\!\!\!\transker\,\exp\left(-i\frac{v\,p}{\hbar}\right) \,dv$, is $\mathcal{T}_{\hbar}(q,p)=T(q,p)$ for linear systems, and $\mathcal{T}_{\hbar}(q,p)=T(q,p)+\mcal{O}(\hbar^2)$ for non-linear systems, where $T(q,p)$ is the LTOA and $\mcal{O}(\hbar^2)$ is the leading quantum correction to the LTOA for non-linear systems. Weyl-quantizing $\mathcal{T}_{\hbar}(q,p)$ gives us an operator $\opr{T}$, the projection of which in the Hilbert space $\mcal{H}_{l}$ gives the above confined quantum time of arrival operators. The CTOA-operators for non-linear systems then is a quantization of the corresponding local time of arrival plus quantum corrections to the classical-LTOA. The observables of $\hat{\mathcal{C}}_O^{(c)}$ then have the correct classical limit.

At this point we have completely characterized the CTOA-operators. For systems that equation-(\ref{kerneldiff}) have solutions with the required properties, e.g. for infinitely differentiable potentials \cite{herb}, the corresponding CTOA-operators are compact and self-adjoint, possessing a complete set of eigenfunctions and a discrete spectrum. But do these eigenfunctions unitarily arrive at the origin at their respective eigenvalues as we have conceived them to be? We refer to the harmonic oscillator for insight. By symmetry arguments, combined with a numerical computation of the first few eigenfunctions, indicate that for a given $\gamma$ and a given positive integer $n$ there is a pair of eigenfunctions $\varphi_{\gamma,n}^{\pm}(q)$ with equal magnitudes of eigenvalues and of opposite signs, i.e. $\tau_{n,\gamma}^+=-\tau_{n,\gamma}^{-}$, with the sign indicating the sign of the eigenvalue. We find that the eigenfunctions can also be classified in the same way as those in \cite{galapon0}; in particular, eigenfunctions can be nodal or non-nodal (see Figure-1). 

The parity eigenfunctions for $\gamma=\frac{\pi}{2}$ 
are numerically determined using Nystrom method employing Gauss-Legendre integration quadrature \cite{delves}, combined with Nystrom's interpolation scheme to get the eigenfunctions at a uniformly spaced grid points. The eigenfunctions are then evolved using Crank-Nicholson differencing of the time-dependent Schrodinger equation. In those cases where the combined numerical computation for the eigenfunctions and the evolution are known to converge, the eigenfunctions are found to unitarily arrive at the origin at their respective eigenvalues within numerical accuracy
(See Figure-1). Moreover, the minimum variance of the eigenfunctions at the eigenvalue decreases with increasing $n$ so that the eigenfunctions become increasingly localized at the origin at their eigenvalues as those in \cite{galapon0}. 
We mention that the same numerical simulation for $\gamma=\frac{\pi}{2}$ has been done for the linear potential, $V=\lambda q$, and the same dynamical behaviors have been observed for the eigenfunctions.

\begin{figure}[!tbp]
{\includegraphics[height=1.3in,width=1.6in,angle=0]{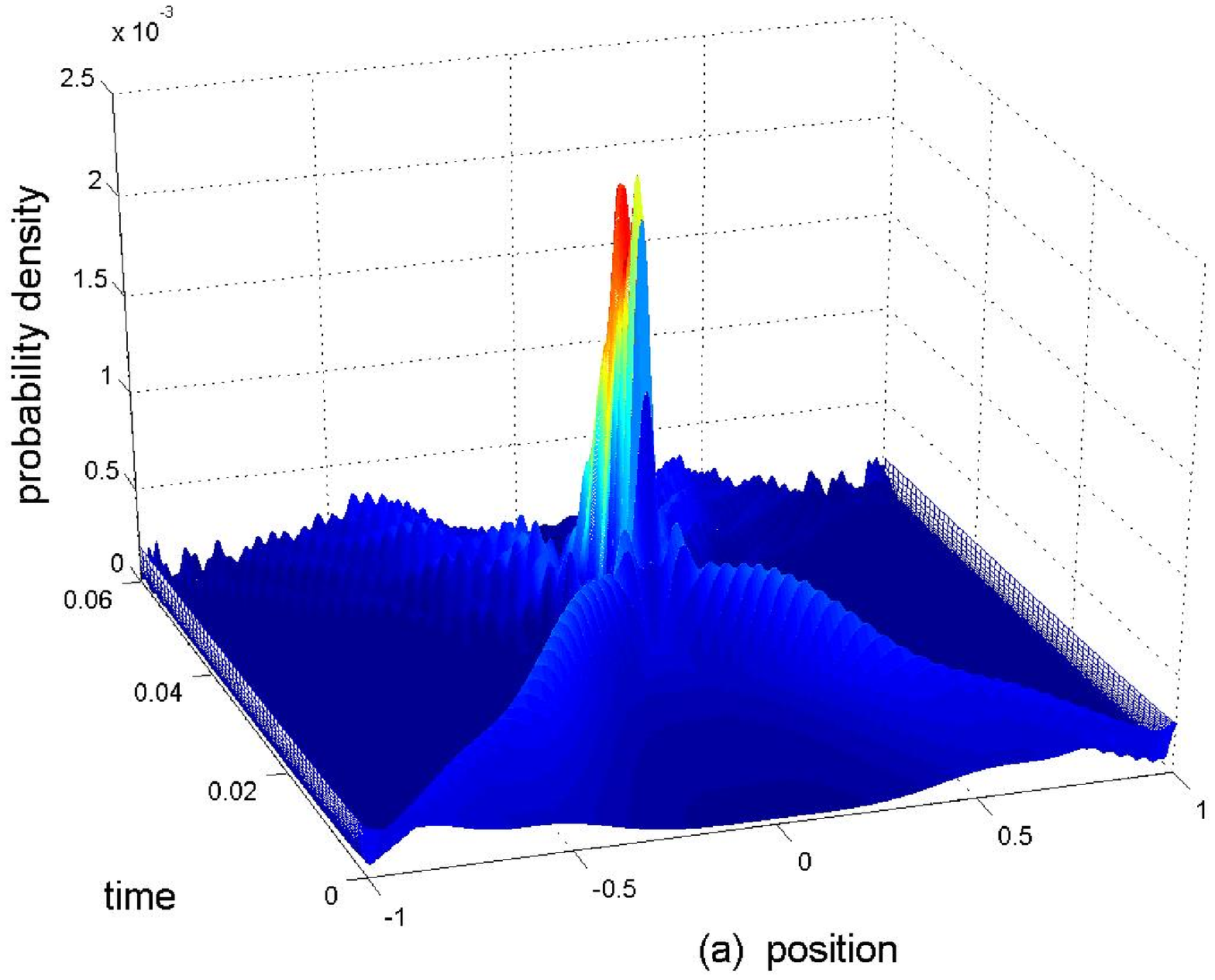}}
{\includegraphics[height=1.3in,width=1.6in,angle=0]{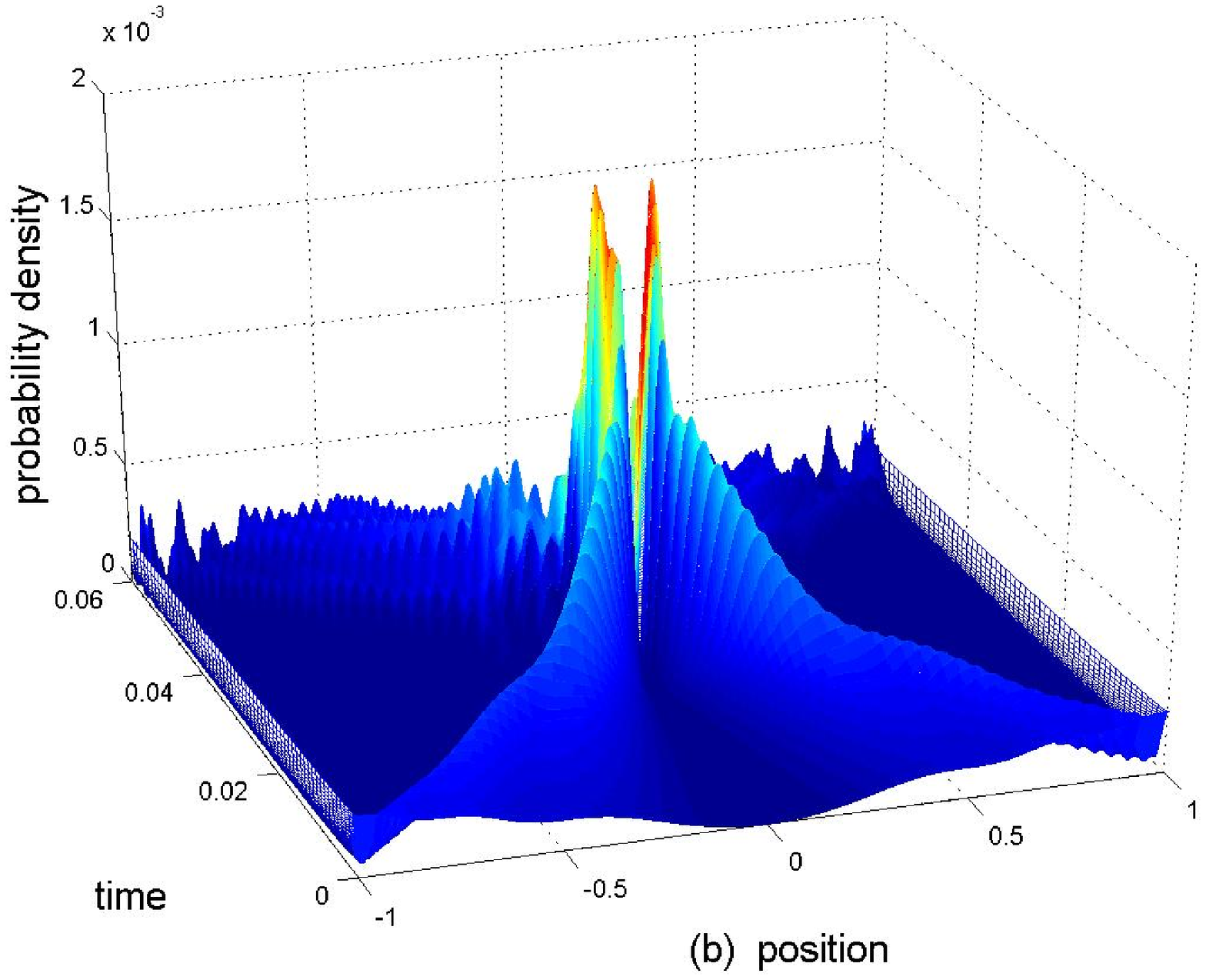}}
{\includegraphics[height=1in,width=3in,angle=0]{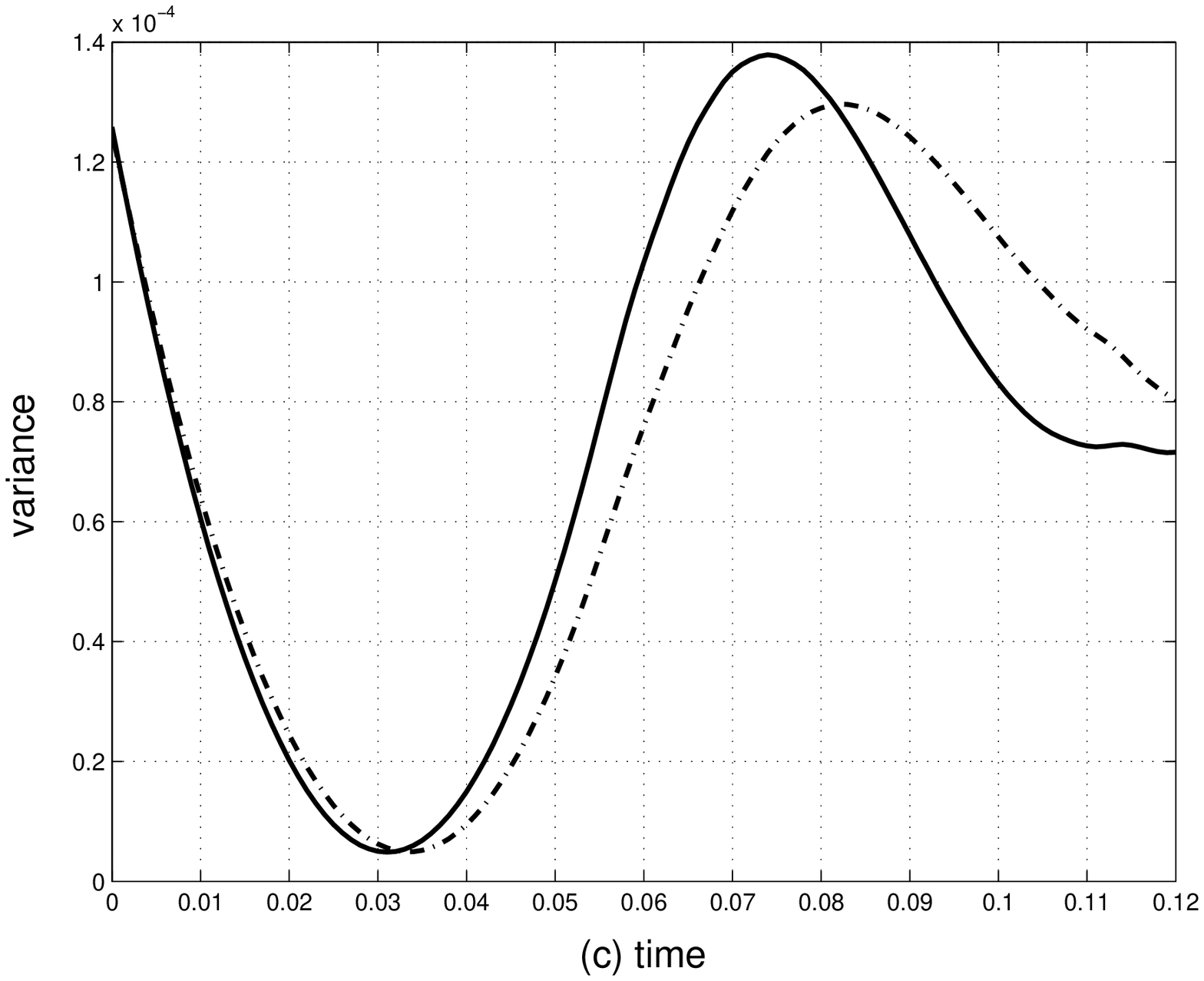}}
\caption{The evolved probability densities corresponding to the eigenfunctions ({\it a}) $n=5$ (non-nodal eigenfunction) and ({\it b}) $n=6$ (nodal eigenfunction) for $\gamma=\frac{\pi}{2}$, $\hbar=l=m=\omega=1$. Both unitarilly arrive at the origin at their respective eigenvalues, 0.0336 and 0.0303. The variances of the position operator is minimum at their eigenvalues as demonstrated by ({\it c}), with dashed line $n=5$ and solid line $n=6$.}
\end{figure}

While more works have to done to completely understand the physical contents of the CTOA-operators, the above numerical results already,  strongly endorse the interpretation of the confined quantum time of arrivals as first time of arrival operators, with the eigenvalues as the first time of arrivals at the origin of their respective eigenfunctions.

The author is supported by the University of the Philippines System through the U.P. Creative and Research Scholarship Program. The author acknowledges the invaluable help of F. Delgado on the numerical aspect of this work, the critical reading of A. Ruschhaupt of this paper, and the insightful discussions with I. Egusquiza and J.G. Muga. This work is partially supported by the ``Ministerio de Ciencia y Tecnolog{\'\i}a'' and FEDER (Grant BFM2003-01003).

\end{document}